\documentclass[10pt,a4paper,aps,pra,twocolumn,showpacs,floatfix]{revtex4}

\usepackage{graphicx}
\usepackage{amsmath,amsthm,amsfonts}
\usepackage{psfrag}
\usepackage{fullpage}
\usepackage{textcomp}
\usepackage{tabularx}
\usepackage{booktabs}
\usepackage{array}
\usepackage{bbm}
\usepackage{bm}
\usepackage{xcolor}

\newcommand{\comment}[1]{}
\newcommand{\hide}[1]{}

\renewcommand{\phi}{\varphi}

\begin{document}
%
\title{Decoherence Effects on the Non-locality of Symmetric States}
\author{Adel Sohbi} \email{adel.sohbi@telecom-paristech.fr}
\author{Isabelle Zaquine}
\author{Eleni Diamanti}
\author{Damian Markham}
\affiliation{CNRS LTCI, Departement Informatique et Reseaux, Telecom ParisTech,\\
23 Avenue d'Italie, CS 51327, 75214 Paris CEDEX 13, France}

%
\begin{abstract}

The observation of the non-local properties of multipartite entangled states is of great importance for quantum information protocols. Such properties, however, are fragile and may not be observed in the presence of decoherence exhibited by practical physical systems. In this work, we investigate the robustness of the non-locality of symmetric states experiencing phase and amplitude damping, using suitable Bell inequalities based on an extended version of Hardy's paradox. We derive thresholds for observing non-locality in terms of experimental noise parameters, and demonstrate the importance of the choice of the measurement bases for optimizing the robustness. For $W$ states, in the phase damping case, we show that this choice can lead to a trade-off between obtaining a high violation of the non-local test and optimal robustness thresholds; we also show that in this setting the non-locality of $W$ states is particularly robust for a large number of qubits. Furthermore, we apply our techniques to the discrimination of symmetric states belonging to different entanglement classes, thus illustrating their usefulness for a wide range of practical quantum information applications.


\end{abstract}

\date{\today}
\maketitle

\section{Introduction}

Entanglement refers to the property of a quantum state of many systems to not be decomposable as product states.
It gives rise to the notion of non-locality, whereby spatially separated observers can create correlations in a way impossible to reproduce by the use of shared classical randomness, or equivalently by a local hidden variable (\emph{lhv}) model \cite{EPR:pr35}. In addition to its fundamental interest, non-locality has proven to be a valuable resource for quantum information in many settings, such as communication complexity \cite{BZP:prl04}, randomness amplification \cite{RBG:arxiv13}, device independent quantum key distribution \cite{ABG:prl07}, and other device independent protocols \cite{BGL:prl11}.

Bell inequalities \cite{Bell:physics64} are used as witnesses to test the appearance of non-locality. Mathematically these are bounds on some expression, which is a linear superposition of probabilities of measurement outcomes, found by assuming the existence of local hidden variables. The violation of such expressions proves the presence of non-local correlations, thus providing an experimentally accessible way to detect such correlations between space-like separated systems. Interestingly, the degree of violation of a Bell inequality for a particular state can be linked to its usefulness for the information processing tasks mentioned above.

Possibly the simplest and most used Bell inequality is the Clauser-Horne-Shimony-Holt (CHSH) \cite{CHSH:prl69} inequality, which pertains to the bipartite case. In the last years, however, a variety of Bell inequalities has been developed and examined for the multipartite setting. In \cite{HCS:njp11}, the authors propose a Bell inequality, which is maximally violated by the $W$ states; in fact, the violation reaches its algebraic maximum in the asymptotic limit for an increasing number of qubits in the state. $W$ states belong to a larger set of states, namely permutation symmetric states. Such states present the important practical advantage that their generation has been extensively studied \cite{TZB:prl07,LRS:pra07} and experimentally achieved using photonic \cite{MLF:prl05,KSTSW:prl07} and trapped ion \cite{HHR:nature05} systems. The entanglement of symmetric states has been recently examined \cite{Markham:pra11,AMM:njp10,RM:prl11}, and suitable Bell inequalities have been developed \cite{WM:prl12,WM:pra13}. The latter are based on Hardy's paradox \cite{Hardy:prl93} and use as a main tool the Majorana representation of symmetric states \cite{Majorana:ilnuovo32}, which facilitates the study of the non-local properties of such states and allows to link these properties to the degeneracy occurring in the representation.

The demonstration of non-local features discussed above holds in the ideal case of dealing with pure states. However, in any practical setting, the non-local properties of quantum states can be degraded due to decoherence experienced by physical systems. Decoherence describes the degrading of a quantum system due to the interaction with its environment. Such noise effects become particularly pronounced in many-particle systems because of the complex nature of interactions between all subsystems. Some studies have considered the effect of noise on the non-locality exhibited by multipartite states. It was shown, for instance, that the asymptotic increase in violation of the inequality of \cite{HCS:njp11} with the number of parties is reversed in the presence of decoherence as non-local correlations become increasingly fragile for high number of qubits \cite{LBA:pra11,CB:pra11}.
We see then that indeed sensitivity to noise can drastically affect conclusions concerning non-locality.

Other studies have considered the non-locality of a variety of states, including GHZ, $W$, or graph states, under various noise models, such as depolarization, dephasing or dissipation \cite{JCK:pra06,LPBZ:pra10,CCA:pra12,CAAC:pra}. Furthermore, the effect of decoherence on non-locality has been examined in the context of loophole-free Bell tests \cite{CS:prx12,BC:pra12,PVB:pra12}, while some works have focused on the finite detection efficiency in such tests, illustrating its importance for the implementation of quantum communication tasks \cite{GCB:pra13,KJ:pra13}.

In this work, we study the robustness of the non-locality exhibited by symmetric states in the presence of decoherence in the form of amplitude or phase damping. Our analysis is based on recently developed Bell inequalities for such states \cite{WM:prl12,WM:pra13}, and aims at developing practical criteria for testing the non-local properties of states that can be produced experimentally, in realistic conditions.

The paper is structured as follows. In Section II, we provide the background of our analysis; in particular, we present the quantum states that we will examine, the noise models under consideration, and suitable non-local tests. In Section III, we describe the methods that we use to quantify the effect of decoherence on the observation of non-locality for the states and tests that interest us. In Section IV, the results of our analytical and numerical models are described in particular cases. We analyze and compare the robustness of several symmetric states for different noise models. In Section V, we compare the Bell tests under study and comment on their behavior regarding the type of noise considered. In Section VI, we discuss the choice of the measurement bases as a relevant factor in the robustness of non-locality. Interestingly, we find that the optimum basis, which gives the highest violation, is, in general, not the basis that leads to the highest robustness in the case of phase damping. In Section VII, we investigate the sensitivity of the Bell inequality violation to small changes in the angular settings of the measurement bases.
Finally, in Section VIII, our techniques are applied to the discrimination of Dicke states using a non-local test that is sensitive to degeneracy.

\section{Background}

\subsection{Symmetric states}\label{sec:states}

A permutation symmetric state of $n$ qubits can be written as:
\begin{equation}
\vert\psi\rangle=\sum\limits_{k=0}^{n}c_k\vert S(n,k)\rangle,
\end{equation}
where $|S(n,k)\rangle$ are the Dicke states:
\begin{equation}\label{eq:Snk}
\vert S(n,k)\rangle= {\binom{n}{k}}^{-\frac{1}{2}} \sum\limits_{perm}\vert\underbrace{0\dots0}_{n-k}\underbrace{1\dots1}_{k}\rangle.
\end{equation}
A representation that is particularly useful for symmetric states was introduced by Majorana \cite{Majorana:ilnuovo32}. In this representation, the quantum state $\vert\psi\rangle$ is expressed as a sum of all permutations of tensor products over a set of $n$ qubits $\{\eta_i\}$, $i\in\{1,n\}$,
\begin{equation}
\vert\psi\rangle=K\sum\limits_{perm}\vert\eta_1\dots\eta_n\rangle,
\end{equation}
where $K$ is a normalization factor. These qubits as mapped onto the Bloch sphere are called the Majorana points of $\vert\psi\rangle$, \emph{i.e.},
\begin{equation}
\vert\eta_k\rangle=\cos\left(\frac{\theta_k}{2}\right)\vert 0\rangle + e^{i\varphi_k}\sin\left(\frac{\theta_k}{2}\right)\vert 1\rangle,
\end{equation}
where $\theta_k$ and $\varphi_k$ are the inclination and azimuthal angles, respectively. This gives a convenient geometric representation of symmetric states of $n$ qubits as $n$ points on the surface of a sphere. We note that it is straightforward to switch between the Dicke and Majorana representations through a one-to-one map (see e.g. \cite{BZ,Markham:pra11}).

In addition to Dicke states themselves, other examples of symmetric states that we consider in this work are the $W$ states, $\vert W_n\rangle = \vert S(n,1)\rangle$; the tetrahedron, $\vert T\rangle = (\vert S(4,0)\rangle+ \sqrt{2}\vert S(4,3)\rangle)/\sqrt{3}$; the cube, $\vert C\rangle = (\sqrt{5}\vert S(8,0)\rangle+ \sqrt{14}\vert S(8,4)\rangle + \sqrt{5}\vert S(8,8)\rangle)/2\sqrt{6}$; the octahedron, $\vert O\rangle = (\vert S(6,1)\rangle+ \vert S(6,5)\rangle)/\sqrt{2}$; and the states $\vert 000+\rangle = (2\vert S(4,0)\rangle+ \vert S(4,1)\rangle)/\sqrt{5}$ and $\vert 00++\rangle = (6\vert S(4,0)\rangle+ 6\vert S(4,1)\rangle+ \sqrt{6}\vert S(4,2)\rangle)/\sqrt{78}$.\\

It is important to note that the Majorana points may not all be distinct for some symmetric states. This leads us to define the degeneracy configuration (DC) of a symmetric state as all the numbers of redundancy of all its Majorana points \cite{BKM:prl09}. Then, the degeneracy, $d$, of a state is defined as the highest among those numbers. Interestingly, the DC constitutes an entanglement classification in the sense that each symmetric state belongs to a single DC class, that is, the degeneracy configuration of a state cannot be modified under stochastic local operations and classical communication (SLOCC) \cite{BKM:prl09}. This course grained classification presents an advantage in that it contains a finite number of classes for a state of $n$ qubits compared to the infinite number of classes contained in the SLOCC classification for $n\ge 4$ qubit states.

\subsection{Decoherence models}

The effect of decoherence on a quantum state can been seen as a map that transforms one density matrix to another, and can be described using the operator sum formalism. In this formalism, we can write the noisy version of a state as
\begin{equation}
\rho_{\text{dec}}=\sum_{\vec{k}=(0,\dots,0)}^{(1,\dots,1)}\mathcal{K}_{\vec{k}}\rho{\mathcal{K}_{\vec{k}}}^\dagger,
\label{eq:noisy rho}
\end{equation}
where $\rho$ is the density matrix of the system before the interaction with the environment, and $\mathcal{K}_{\vec{k}}$ is the tensor product of a particular combination of Kraus operators, $K$, given by $\vec{k}$, that is $\mathcal{K}_{\vec{k}} = \otimes_{i}K_{k_i}$.

In this work, we will consider two noise models that are relevant for practical implementations, namely amplitude and phase damping. The former essentially describes losses that are ubiquitous in experimental setups, while phase damping appears, for instance, in photonic systems implementing quantum communication protocols \cite{BCT:njp12}.

For amplitude damping, the Kraus operators are
\begin{equation}
 K_0 =
 \begin{pmatrix}
  0 & \sqrt{\gamma}\\
  0 & 0 \\
 \end{pmatrix}
; K_1 =
 \begin{pmatrix}
  1 & 0 \\
  0 & \sqrt{1-\gamma}\\
 \end{pmatrix},
\label{eq:Krausamp}
\end{equation}
where the coefficient $\gamma$ can be interpreted as the probability of losing a photon. For phase damping, which can also been seen as a phase flip channel, they are given by:
\begin{equation}
 K_0 =
 \begin{pmatrix}
  1 & 0 \\
  0 & \sqrt{1-\lambda}\\
 \end{pmatrix}
; K_1 =
 \begin{pmatrix}
  0 & 0 \\
  0 & \sqrt{\lambda}\\
 \end{pmatrix},
\label{eq:Krausphase}
\end{equation}
where here $\lambda$ can be interpreted as the probability for a photon to be scattered. Further details on these noise models can be found in \cite{LOM:pra04,NielsenChuang}.

\subsection{Non-local tests for symmetric states}

The main test of non-locality that we will consider in this work is based on an extended version of Hardy's paradox. The original paradox is a two-party logical proof of non-locality \cite{Hardy:prl93}, and is defined as a set of four probabilistic conditions. When one takes all the conditions together, this leads to a logical contradiction. The extended Hardy's paradox can be seen as a $n$-party game: $n$ parties use the measurement settings `0' or `1' and may obtain two outcomes, `0' or `1', for each setting. Compared to the original Hardy's paradox, the number of conditions increases as a consequence of the increased number of parties. The first condition imposes that if all parties use the setting `0' there is a non zero probability that they all obtain the outcome `0':
$$P(0\dots0|0\dots0)>0.$$
Here we denote the probability of getting outcomes $\vec{r}$ for settings $\vec{M}$ as $P(\vec{r}|\vec{M})$.
The next condition (which is actually a set of conditions) is similar to the first one, but one of the parties uses the setting `1'. In this case the joint probability for all of them to obtain the outcome `0' is zero:
\begin{align*}
P(0\dots0|10\dots0) & = 0,\\
P(0\dots0|01\dots0) & = 0,\\
\dots & \\
P(0\dots0|0\dots01) & =0.
\end{align*}
These conditions can be gathered into one expression:
$$\sum_{\pi}P(0\dots0|\pi(0\dots01))=0,$$
where $\pi\in S_n$ denotes the group of permutations of $n$ objects, $S_n$. The final condition states that if all parties use the setting `1', the probability to obtain the outcome `1' for all the parties is zero:
$$P(1\dots1|1\dots1)=0.$$

According to a \emph{lhv} model, a joint probability of obtaining the outcomes $r_i$ with the settings $M_i$ is described by the following expression:
\begin{align}
P(r_1,...,r_n|M_1,...,M_n) & = \notag\\
                           & \int\limits_{\Lambda}q(\lambda)\prod\limits_{i=1}^{n}P(r_i|M_i,\lambda)d\lambda,
\label{eq:lhv}
\end{align}
where $q(\lambda)$ is the probability distribution of the hidden variable $\lambda$ in the space $\Lambda$. It is then possible to show that if a system verifies all the above conditions this leads to a logical contradiction, hence proving that such a system cannot have a possible \emph{lhv} description.

Based on the above analysis, it is possible to derive the following Bell inequality \cite{WM:prl12}:
\begin{align}\label{eq:Pn}
\mathcal{P}^n & := P(0\dots0|0\dots0) - P(1\dots1|1\dots1)\notag\\
							&- \sum_{\pi} P(0\dots0|\pi(0\dots01)\leq 0.
\end{align}
Using the Majorana representation for permutation symmetric states, it was proven that all such states violate the above inequality, \emph{i.e.}, they satisfy $\mathcal{P}^n > 0$ \cite{WM:prl12}, while more recently this was also shown for all pure states \cite{YCZ:prl12}.\\

Based on similar techniques, the extended version of Hardy's paradox can also be used to construct a Bell inequality that is sensitive to the degeneracy of a symmetric state \cite{WM:prl12}:
\begin{align}\label{eq:Qnd}
\mathcal{Q}^n_d & := \mathcal{P}^n - P(\underbrace{1\dots1}_{n-1}|\underbrace{1\dots1}_{n-1}) - \dots \notag\\
                & - P(\underbrace{1\dots1}_{n-d+1}|\underbrace{1\dots1}_{n-d+1})\leq0.
\end{align}
Note that in this expression, since the state is symmetric, a probability concerning only a subspace of the state is independent of the space that is traced out. It was shown that any state with degeneracy $d$ will violate $\mathcal{Q}_d^n$ \cite{WM:prl12}, hence illustrating that the degeneracy of symmetric states can indeed be detected using their non-local properties.\\

In addition to the above inequalities, we will consider for comparison purposes an extended version of the inequality developed in \cite{HCS:njp11}, tailored to high order Dicke states. This inequality is given by the expression \cite{WM:pra13}:
\begin{align}\label{eq:Hnk}
\mathcal{H}^n_k & := \sum_{\pi} P(\pi(\underbrace{0\dots0}_{n-k}\underbrace{1\dots1}_k)|0\dots0)\notag\\
                & - \sum_{\pi} P(\pi(\underbrace{0\dots0}_{n-k-1}\underbrace{1\dots1}_{k-1}01)|\pi(\underbrace{0\dots0}_{n-2}11))\notag\\
                & - P(0\dots0|1\dots1) - P(1\dots1|1\dots1) \leq 0.
\end{align}

\section{Methods}\label{sec:methods}

We begin our analysis by detailing our approach for quantifying the effect of noise on the non-locality of symmetric states. First, we use Eqs. (\ref{eq:noisy rho}), (\ref{eq:Krausamp}), and (\ref{eq:Krausphase}), to calculate the noisy version of the state under study, $\rho_{\text{dec}}$, for both noise models. To this end, we sum over all combinations of Kraus operator elements describing amplitude and phase damping applied to the pure density matrix of the system. This allows us to calculate the \emph{fidelity} of the noisy state with respect to the initial pure state, which we will denote in the following as $F_{\text{amp}}$ and $F_{\text{ph}}$ for the two noise models, respectively. These are functions of the corresponding coefficients of the Kraus operators, $\gamma$ and $\lambda$, respectively. We call these coefficients the \emph{noise factors}. For the amplitude damping case, we additionally ``translate'' the probability of absorption, $\gamma$, to a detection efficiency, assuming that the probability to lose a photon is mostly due to the detection process in a practical experimental setup. In particular, we define the detection efficiency as $\eta = \sqrt{1-\gamma}$, and attribute a different detection efficiency, $\eta_0$ and $\eta_1$, to the two measurement settings of our non-local tests, `0' and `1', respectively. This choice corresponds to typical scenarios of interest in photonics experiments.

The second step is to compute the probabilities in the expressions corresponding to the inequalities of Eqs. (\ref{eq:Pn}), (\ref{eq:Qnd}), and (\ref{eq:Hnk}), and hence to derive the degree of violation achieved by the noisy states in each case. For this purpose, it is necessary to choose a measurement strategy, \emph{i.e}, a measurement basis for each setting. For simplicity, we assume that all parties make the same basis choice. This may not be optimal in general, but allows for the numerical and analytical solutions found here, and gives interesting bounds on robustness. We will consider in the following different possible choices: in particular, the parties can measure using the \emph{Majorana basis}, which is defined by the Majorana points of the pure state through the procedure outlined in \cite{WM:prl12}, or the \emph{optimum basis}, where optimum here refers to the fact that the specific basis choice leads to a maximal violation of the corresponding Bell inequality for the pure state. Once the measurement strategy has been decided, then, assuming projective measurements, the probabilities can be calculated as follows:
\begin{equation}
P(r_1\dots r_n|M_1\dots M_n) = tr\lbrace\rho_{\text{dec}}\bigotimes_{i=1}^n \Pi_{r_i|M_i}\rbrace,
\end{equation}
where $r_i, M_i$ denote the outcomes and measurement settings, respectively, for qubit $i$. The $\Pi_{r_i|M_i}$ are projectors on the Bloch sphere and can be written as $\Pi_{r_i|M_i}=|b_{r_i|M_i}\rangle\langle b_{r_i|M_i}|$, with $|b_{r_i|M_i}\rangle = \cos\left(\dfrac{\theta_{M_i}}{2}-r_i\dfrac{\pi}{2}\right)|0\rangle+e^{i\phi_{M_i}}\sin\left(\dfrac{\theta_{M_i}}{2}-r_i\dfrac{\pi}{2}\right)
|1\rangle,$ where $\theta_{M_i}$ and $\varphi_{M_i}$ are the inclination and azimuthal angles on the Bloch sphere, respectively. In this way, $\mathcal{P}^n$, $\mathcal{Q}^n_d$, and $\mathcal{H}^n_k$ can be written as polynomial functions of the angles of the measurement bases and the noise factors only.

The final step is then to determine suitable thresholds for non-locality, which are derived by setting the obtained expressions for $\mathcal{P}^n$, $\mathcal{Q}^n_d$, and $\mathcal{H}^n_k$ to zero. For the amplitude damping case, these thresholds can be expressed either by the values of the fidelity, $F_{\text{amp,th}}$, or the detection efficiencies, $\eta_{0,\text{th}}$ and $\eta_{1,\text{th}}$, below which it is not possible to prove non-locality for a given Bell test and a given measurement strategy, or by the value of the noise factor, $\gamma_{\text{th}}$, above which again a violation cannot be achieved. The noise factor and fidelity thresholds are calculated assuming $\eta_0 = \eta_1$. For the phase damping case, the corresponding thresholds refer either to the noise factor, $\lambda_{\text{th}}$, or to the fidelity, $F_{\text{ph,th}}$. These various criteria for characterizing the robustness will be useful for comparing different states, measurement strategies, and non-local tests under realistic conditions of interest.

\section{Results for the $\mathcal{P}^n$ test}\label{sec:results}

We apply the method described previously to quantify the non-local properties of several symmetric states under decoherence based on the Bell inequality of Eq. (\ref{eq:Pn}). Our goal is to determine whether it is possible to observe such properties in realistic environments.

\subsection{Examples}

We start by presenting our results first for Dicke states and then for two specific examples of symmetric states, namely the $W$ states and the tetrahedron state. In each case, our results correspond to a specific choice of a measurement strategy. The role of this choice will be discussed in Section \ref{sec:measurement}.

\subsubsection{Dicke states}

Starting from the pure Dicke states, Eq. (\ref{eq:Snk}), we can write the states under phase damping noise as follows:
\begin{equation}
\rho_{\text{dec,ph}}=\sum_{\vec{k}=(0,\dots,0)}^{(1,\dots,1)}\mathcal{K}_{\vec{k}}|S(n,k)\rangle\langle S(n,k)|{\mathcal{K}_{\vec{k}}}^\dagger,
\end{equation}
where $\mathcal{K}_{\vec{k}}=\bigotimes_{i=1}^{n}K_{k_i}$, $k_i\in\{0,1\}$, and $K_0, K_1$ are given in Eq. (\ref{eq:Krausphase}). Since Bell tests are linear, we can then write an analytical expression for the $\mathcal{P}^n$ value of this state as a superposition of the values corresponding to each component in the state. Using the Majorana measurement bases defined by the two settings $\mathcal{M}_0 =\{\theta_0=\pi/2$, $\varphi_0=0\}$ and $\mathcal{M}_1 =\{\theta_1=\pi$, $\varphi_1=\pi\}$ for all parties, we find:
\begin{widetext}
\begin{align}\label{eq:PnSnkph}
\mathcal{P}^n(\rho_{\text{dec,ph}})& = \mathcal{P}^n(K_0^{\otimes n}|S(n,k)\rangle\langle S(n,k)| K_0^{\otimes n})
+\sum_{j=1}^{k}\sum_{perm}\mathcal{P}^n(K_0^{\otimes n-j}\otimes K_1^{\otimes j}|S(n,k)\rangle\times h.c.)\notag\\
& = \frac{1}{2^n}\left(\binom{n}{k}(1-\lambda)^k(1-\frac{2k^2}{n})+\sum_{j=1}^{k}\lambda^j(1-\lambda)^{k-j}\binom{k}{j}\binom{n-j}{k-j}(1-2k)\right).
\end{align}

Following a similar procedure for the amplitude damping case, we find:
\begin{gather}\label{eq:PnSnkamp}
\mathcal{P}^n(\rho_{\text{dec,amp}})=\frac{1}{2^n}\left(\binom{n}{k}(1-\gamma)^k(1-\frac{2k^2}{n})+
\sum_{j=0}^{k}\gamma^j(1-\gamma)^{k-j}\binom{n}{j}\binom{n}{k}^{-1}\binom{n-j}{k-j}(1-2(k-j))\right)-\gamma^k.
\end{gather}
\end{widetext}

We first remark that, in the absence of noise, a violation of $\mathcal{P}^n$ can be obtained only under the condition that $k \leq\sqrt{n/2}$. This means that for large $k$, a violation can only be observed for a high number of qubits in the state.

The fidelities between the pure Dicke states and the states that have experienced decoherence are given by the following expressions:
\begin{equation}
F_{\text{ph}} = \left((1-\lambda)^k+\sum_{j=1}^{k}\binom{n}{j}\binom{n}{k}^{-2}\binom{n-k}{k-j}^2\right)^{1/2}
\end{equation}
\begin{equation}
F_{\text{amp}} = (1-\gamma)^{k/2}.
\end{equation}

\begin{figure}[tb]
      \centering
      \includegraphics[width=8cm]{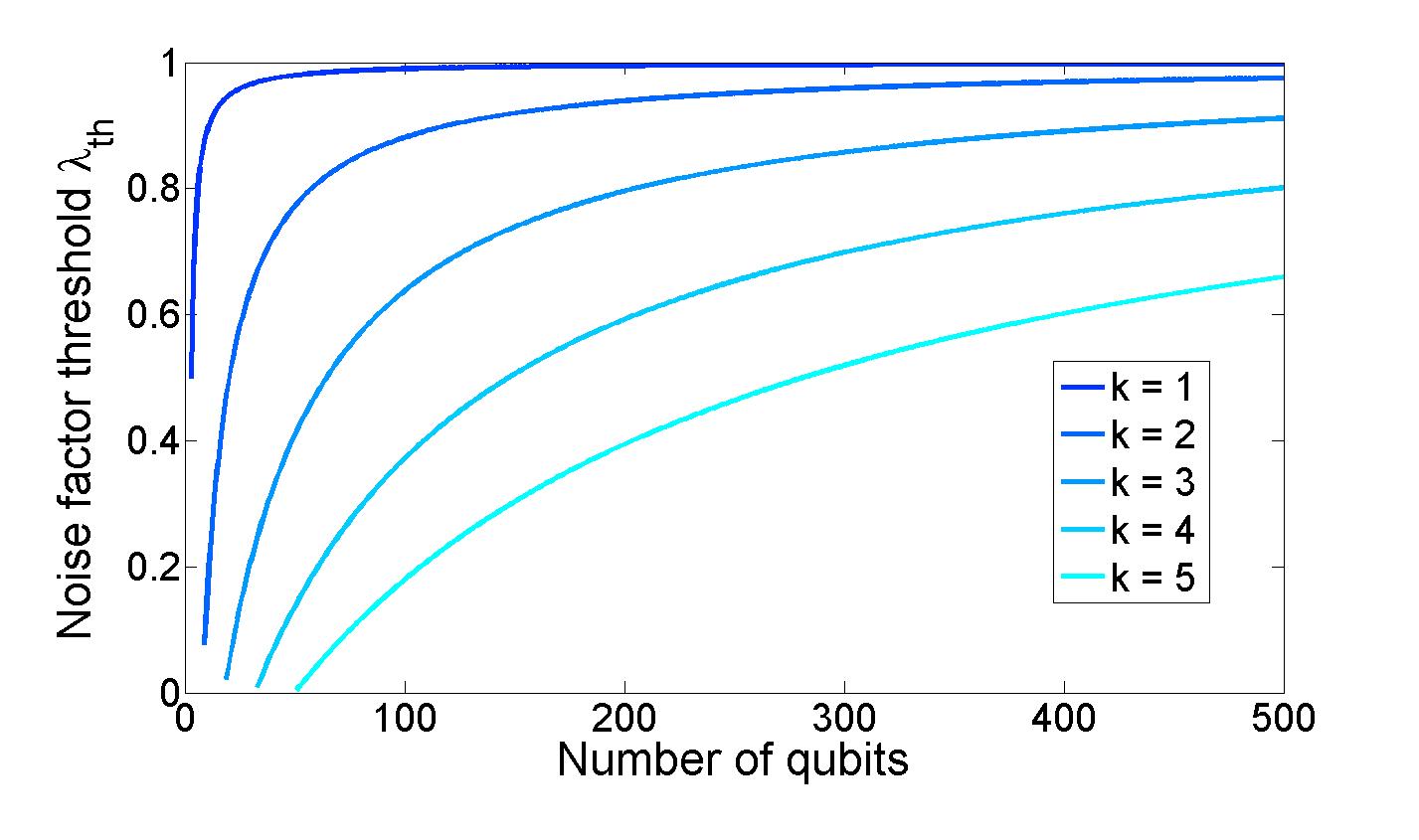}
      \caption{Noise factor threshold as a function of number of qubits for different Dicke states under phase damping noise.}
			\label{fig:SNk_ph}
\end{figure}

\begin{figure}[tb]
      \centering
      \includegraphics[width=8cm]{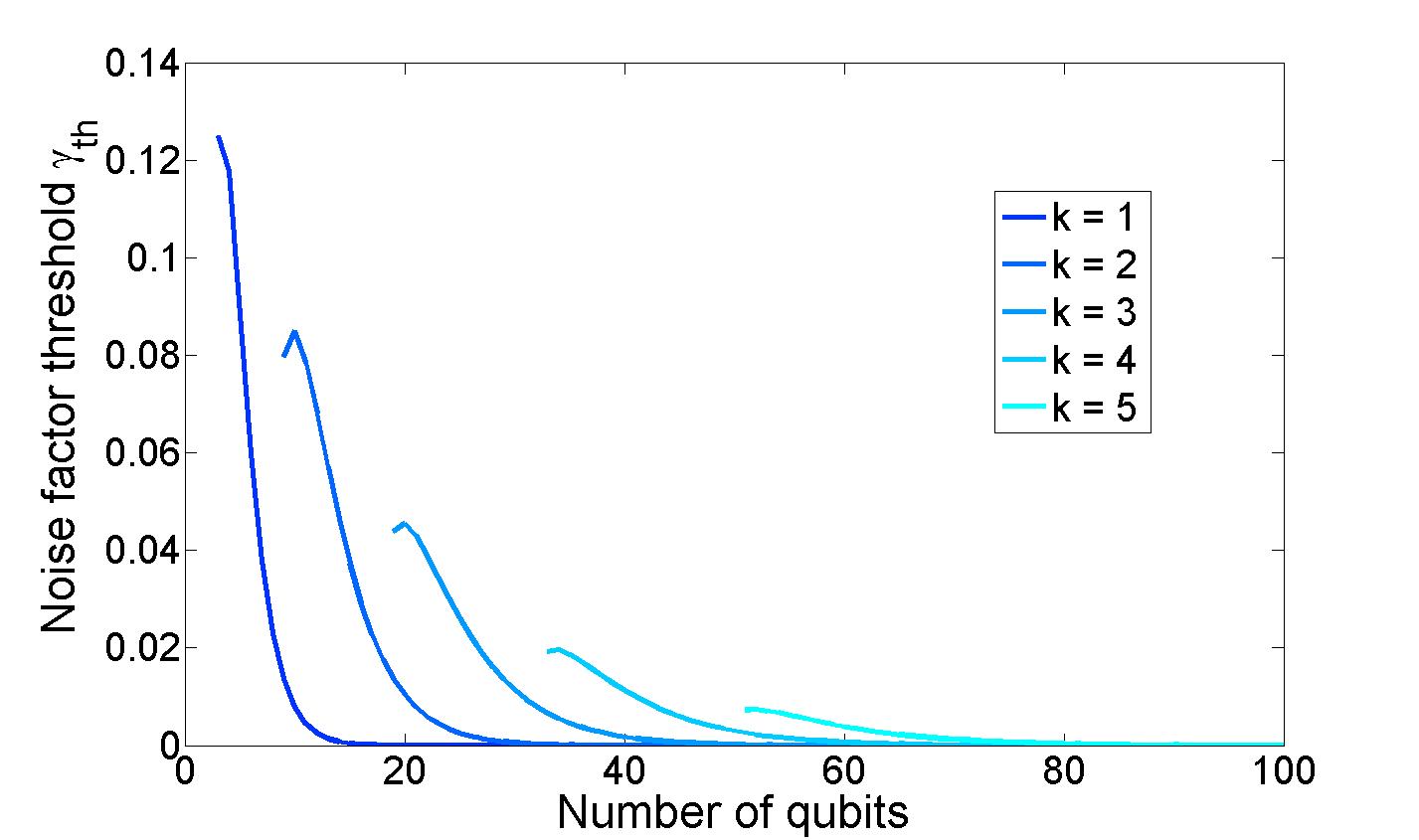}
      \caption{Noise factor threshold as a function of number of qubits for different Dicke states under amplitude damping noise.}
			\label{fig:SNk_amp}
\end{figure}

By setting the expressions of Eqs. (\ref{eq:PnSnkph}) and (\ref{eq:PnSnkamp}) to zero, we can calculate the noise factor thresholds, $\lambda_{\text{th}}$ and $\gamma_{\text{th}}$, as a function of the number of qubits in the Dicke state, $n$, for various $k$. The results are shown in Figs. \ref{fig:SNk_ph} and \ref{fig:SNk_amp} for phase and amplitude damping, respectively. It is interesting to note that the behavior of the noise factor thresholds is very different for the two types of noise. In the phase damping case, the threshold increases with the total number of qubits, but decreases with increasing $k$, while in the amplitude damping case, the threshold decreases rapidly to zero. This shows that states with small $k$ (respecting the condition $k\leq\sqrt{n/2}$) can withstand more phase damping noise, while all states are quite sensitive to amplitude damping noise. In general, the robustness of the Dicke states depends crucially on the type of noise; this is true for all symmetric states and will be further discussed in the following.


\subsubsection{$W$ states}\label{sec:W}

The $W$ states are a special case of the Dicke states, corresponding to $k = 1$. Whereas the solutions that we found for Dicke states are numerical, for the $W$ states we derive analytical bounds. Their Dicke representation is very simple:
\begin{equation}
\vert W_n\rangle = \vert S(n,1)\rangle
\end{equation}
In terms of Majorana representation, these states have one Majorana point corresponding to the state $\vert 1\rangle$, while all the others correspond to $\vert 0\rangle$. The Majorana measurement bases are then defined as $\mathcal{M}_0 =\{\theta_0=\pi/2$, $\varphi_0=0\}$ and $\mathcal{M}_1 =\{\theta_1=\pi$, $\varphi_1=\pi\}$. Note that because of the state geometry, a measurement is invariant under any rotation on the plane formed by the eigenvectors of the Pauli matrices $\sigma_x$ and $\sigma_y$. Assuming all parties use the above measurement settings, we find that a violation of $\mathcal{P}^n$ can be observed for the pure states for all $n$.

For the amplitude damping case, these violations are shown in Table I, for the states $W_{3,4,5,6}$, together with the detection efficiency thresholds, $\eta_{0,\text{th}}$ and $\eta_{1,\text{th}}$, corresponding to the two measurement settings.

\begin{table}[h!]
\center
\begin{tabular}{|l||c|c|c|}
  \hline
  State & $\mathcal{P}^n$ & $\eta_{0,\text{th}}$ & $\eta_{1,\text{th}}$ \\
  \hline
   $W_3$ & 0.1250 & 70.7$\%$ & 91.3$\%$ \\
   $W_4$ & 0.1250 & 57.7$\%$ & 92.6$\%$ \\
   $W_5$ & 0.0938 & 50.5$\%$ & 94.8$\%$ \\
   $W_6$ & 0.0625 & 45.8$\%$ & 96.7$\%$ \\
   \hline
\end{tabular}
   \caption{$\mathcal{P}^n$ values for the pure states $W_{3,4,5,6}$ and corresponding detection efficiency thresholds for amplitude damping noise.}
\end{table}

An interesting observation here is that there is an asymmetry between the detection efficiency thresholds for the two settings. This is due to the structure of $\mathcal{P}^n$, which is not symmetric with respect to the settings `0' and `1' (see Eq. (\ref{eq:Pn})). We also note that $\eta_{0,\text{th}}$ decreases with the number of qubits in the $W$ state, while the opposite is true for $\eta_{1,\text{th}}$. Furthermore, in general, the $W$ states under study can withstand more losses in the `0' than in the `1' setting.

If we set equal detection efficiencies in the two settings, \emph{i.e.,} for $\eta_0 = \eta_1$, we can derive analytical expressions for the noise factor and fidelity thresholds, as follows:
\begin{equation}
\gamma_{\text{th}}= \frac{n-2}{2^{n}+n-3}, \mbox{ }F_{\text{amp,th}} = \sqrt{\frac{2^{n}-1}{2^{n}+n-3}},
\end{equation}
again for measuring in the Majorana bases. The fidelity threshold tends to 1 for a large number of qubits, which means that it becomes hard to demonstrate non-locality in practice in this case.

\begin{figure}[tb]
      \centering
      \includegraphics[width=7cm]{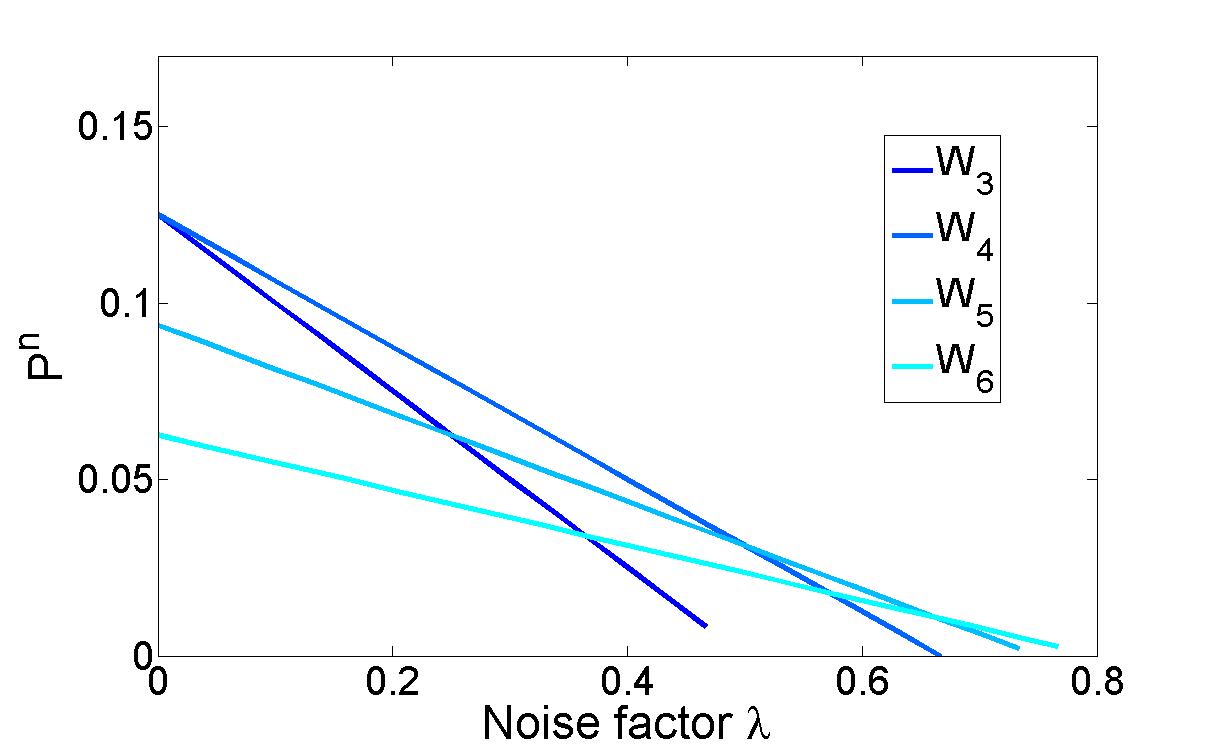}
      \caption{$\mathcal{P}^n$ values as a function of the noise factor for the $W_{3,4,5,6}$ states under phase damping noise.}
			\label{fig:Wstate_ph}
\end{figure}

For the phase damping case, using the same measurement strategy as before, we calculate the $\mathcal{P}^n$ values as a function of the noise factor, $\lambda$, for the $W_{3,4,5,6}$ states. Note that we use the same bases for all states, adding a Majorana point corresponding to $\vert 0 \rangle$ for each added qubit. The results are shown in Fig. \ref{fig:Wstate_ph}. Interestingly, the violation decreases but the curves become flatter with an increasing number of parties. This means that a larger noise can be tolerated for higher $n$, hence leading to an increased robustness, albeit at the expense of a smaller violation.

It is possible to derive in this case as well analytical expressions for the noise factor and fidelity thresholds:
\begin{equation}\label{eq:Wstate_th}
\lambda_{\text{th}} = \frac{n-2}{n-1}, \mbox{ }F_{\text{ph,th}} = \sqrt{\dfrac{2}{n}}.
\end{equation}
Contrary to the amplitude damping case, here the fidelity threshold decreases with an increasing number of qubits, indicating a particularly good robustness obtained in this case. We remark again that, as with general Dicke states, the robustness of $W$ states to phase damping noise is better than the one to amplitude damping. Note also that the robustness of $W$ states to dephasing has been previously reported in the literature \cite{LPBZ:pra10,CAAC:pra}.

\subsubsection{Tetrahedron state}

The last example that we will consider in detail is the tetrahedron \cite{Markham:pra11}, a 4-qubit state with the following Dicke representation \cite{Aulbach:PhD11}:
\begin{equation}
\vert T\rangle = \sqrt{\frac{1}{3}}\vert S(4,0)\rangle+ \sqrt{\frac{2}{3}}\vert S(4,3)\rangle.
\end{equation}
In terms of Majorana representation, this state has the Majorana points: $\vert \eta_1\rangle=\vert 0\rangle$ and $\vert\eta_{2,3,4}\rangle=\sqrt{\frac{1}{3}}\vert 0\rangle+ e^{i\xi}\sqrt{\dfrac{2}{3}}\vert 1\rangle$, where $\xi=\dfrac{\pi}{3},\pi,\dfrac{5\pi}{3}$. Because of its geometry, each combination of measurement bases can be reproduced four different times (corresponding to the four different vertices of the tetrahedron). For the pure tetrahedron state a violation of $\mathcal{P}^n = 0.1621$ is obtained with the Majorana measurement bases defined by $\mathcal{M}_0 = \{\theta_0=0.899$, $\varphi_0=2.435\}$ and $\mathcal{M}_1 = \{\theta_1=2.005$, $\varphi_1=4.285\}$. In Fig. 4 and 5 we assume that all parties choose these measurement settings.

\begin{figure}[tb]
      \centering
      \includegraphics[width=8cm]{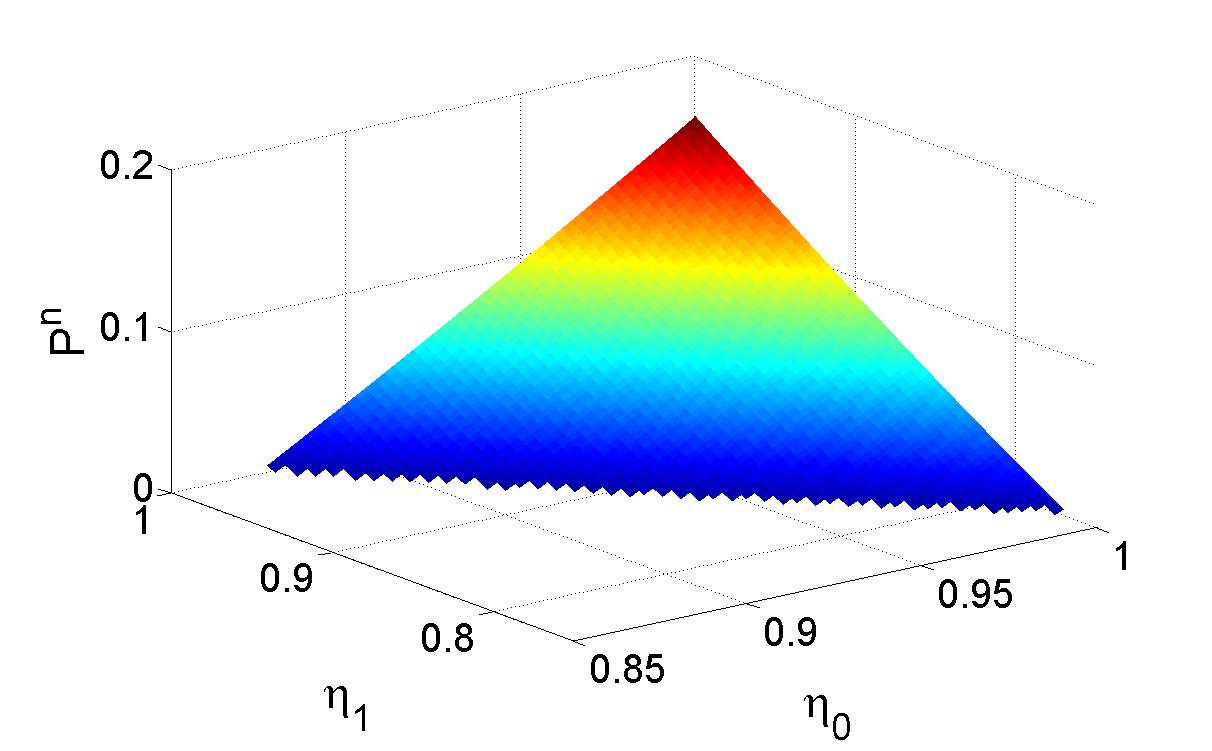}
      \caption{$\mathcal{P}^n$ values  as a function of the detection efficiencies of the two settings for the tetrahedron state under amplitude damping noise.}
			\label{fig:Tstate_Adel_amp}
\end{figure}

For the amplitude damping case, we show in Fig. \ref{fig:Tstate_Adel_amp} the values of $\mathcal{P}^n$ as a function of the detection efficiencies, $\eta_0$ and $\eta_1$, corresponding to the two measurement settings defined above. These results allow us to derive the detection efficiency thresholds; we find $\eta_{0,\text{th}} = 87.18\%$ and $\eta_{1,\text{th}} = 76.16\%$. It is interesting to note that for this choice of measurement strategy, more losses can be tolerated in the `1' than `0' in the setting, contrary to what we found for the $W$ states. However, for the optimum basis, which is defined by the settings $\mathcal{M}_0 = \{\theta_0=1.885$, $\varphi_0=1.047\}$ and $\mathcal{M}_1 = \{\theta_1=0.105$, $\varphi_1=4.189\}$, and leads to the maximal violation $\mathcal{P}^n = 0.1638$, we find that the situation is inverted, with $\eta_{0,\text{th}} = 71.41\%$ and $\eta_{1,\text{th}} = 90\%$.  This illustrates the importance of the choice of the measurement strategy for demonstrating non-locality in practice.


\begin{figure}[tb]
      \centering
      \includegraphics[width=8cm]{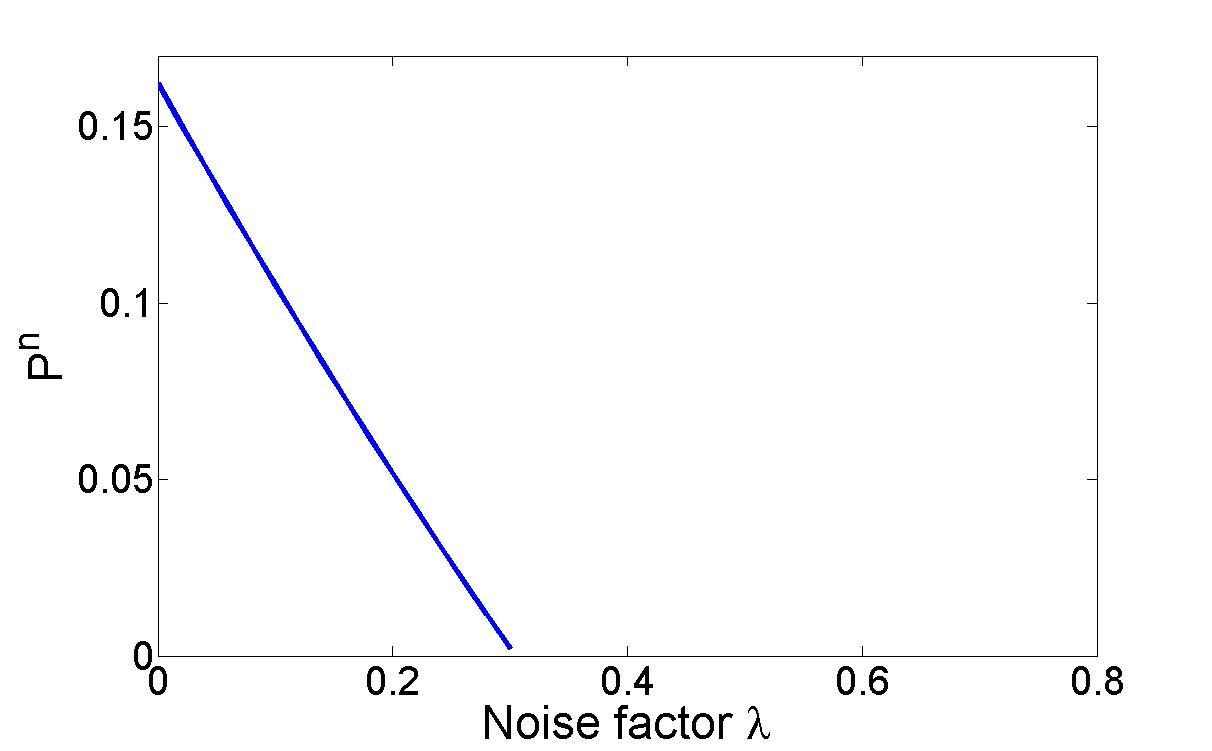}
      \caption{$\mathcal{P}^n$ values as a function of the noise factor for the tetrahedron state under phase damping noise.}
			\label{fig:Tstate_Adel_ph}
\end{figure}

For the phase damping case, we show in Fig. \ref{fig:Tstate_Adel_ph} the values of $\mathcal{P}^n$ as a function of the noise factor, $\lambda$, with the same measurement bases as in the amplitude damping case. The noise factor threshold takes the value $\lambda_{\text{th}}=0.3$.

\subsection{Comparison between symmetric states}

The detailed analysis of a representative set of symmetric states has allowed us to identify some important features concerning the behavior of the non-local properties of such states in the presence of decoherence. We will now provide a comprehensive comparison of the robustness of this set of states, complemented with some additional states, based on the detection efficiency and fidelity thresholds defined previously, pertaining to the Hardy paradox Bell test of Eq. (\ref{eq:Pn}).

In Table II, we summarize our results for several symmetric states containing a variable number of qubits: we show the $\mathcal{P}^n$ values of the pure states obtained in each case with the measurement strategy that we have called optimum because it maximizes the violation achieved by the state, as well as the detection efficiency thresholds, $\eta_{0,\text{th}}$ and $\eta_{1,\text{th}}$, obtained for the amplitude damping noise case as explained in Section \ref{sec:methods}. Below these thresholds, it is not possible to observe a violation of the non-local test for the state under study.

\begin{table}[tb]
\center
\begin{tabular}{|l||c|c|c|}
  \hline
  State & $\mathcal{P}^n$ & $\eta_{0,\text{th}}$ & $\eta_{1,\text{th}}$ \\
  \hline
   \multicolumn{4}{|c|}{3 qubits} \\
	\hline
	$W_3$ & 0.1926 & 64.07$\%$ & 85.40$\%$ \\
	\hline
  \multicolumn{4}{|c|}{4 qubits} \\
	\hline
	$S(4,2)$ & 0.1407 & 76.81$\%$ & 90$\%$ \\
  $W_4$ & 0.1811 & 64.04$\%$ & 85.36$\%$ \\
  Tetrahedron & 0.1638 & 71.41$\%$ & 90$\%$ \\
  $\vert 000+\rangle$ & 0.0141 & 82.46$\%$ & 73.48$\%$ \\
  $\vert 00++\rangle$ & 0.0194 & 90$\%$ & 90$\%$ \\
	\hline
  \multicolumn{4}{|c|}{5 qubits} \\
	\hline
  $W_5$ & 0.1835 & 55.64$\%$ & 86.80$\%$ \\
  	\hline
  \multicolumn{4}{|c|}{6 qubits} \\
	\hline
  $W_6$ & 0.1815 & 53.08$\%$ & 87.06$\%$ \\
  Octahedron & 0.1234 & 72.80$\%$ & 92.74$\%$ \\
    	\hline
  \multicolumn{4}{|c|}{8 qubits} \\
	\hline
	Cube & 0.0890 & 79.37$\%$ & 79.37$\%$ \\
	$W_8$ & 0.1791 & 49.07$\%$ & 87.46$\%$ \\
  \hline
\end{tabular}
  \caption{$\mathcal{P}^n$ values of the pure states and corresponding detection efficiency thresholds for amplitude damping noise.}
\end{table}

As we have noted previously, the asymmetry in the two detection efficiency thresholds is due to their different role in the construction of the $\mathcal{P}^n$ inequality. In general, most symmetric states under study present a lower threshold in the `0' than in the `1' measurement setting, which means that they can tolerate more losses in the former than in the latter. We also note that the $\vert 000+\rangle$ state, the cube, as well as the $W$ states feature thresholds lower than $90\%$ in both settings, which designates those as the most robust states amongst the ones analyzed.

Another important observation concerns the value of the violation that can be expected in a realistic scenario. Clearly, even when a pure state can achieve a high violation, if the corresponding detection efficiency thresholds are very high, it will actually be difficult to observe non-locality in practice \cite{LBA:pra11,CB:pra11}. It is therefore important to take into account all the relevant parameters, namely the desired violation and the characteristics of the available experimental equipment, when designing a non-local test for quantum information applications. This suggests that it is possible to provide a classification of symmetric states in terms of suitability for a given experimental setup.

In Table III, we provide a comparison of the two types of noise that we have considered by presenting the fidelity thresholds for all symmetric states under study, obtained using the optimum measurement strategy in each case. For amplitude damping, this threshold is derived by assuming the same detection efficiency in both settings. Again, for fidelities below these thresholds, it is not possible to demonstrate non-locality in practice.

\begin{table}[tb]
\center
\begin{tabular}{|l||c|c|c|c|}
  \hline
  State & $\mathcal{P}^n$ & $F_{\text{amp,th}}$ & $F_{\text{ph,th}}$ \\
	\hline
  \multicolumn{4}{|c|}{3 qubits} \\
	\hline
  $W_3$ & 0.1926 & 90.04$\%$& 79.16$\%$\\
    \hline
  \multicolumn{4}{|c|}{4 qubits} \\
  \hline
  $S(4,2)$ & 0.1407 & 86$\%$& 81.34$\%$\\
  $W_4$ & 0.1811 & 89.94$\%$& 77.14$\%$\\
  Tetrahedron & 0.1638 & 85.62$\%$& 77.15$\%$\\
  $\vert 000+\rangle$ & 0.0141 & 99.48$\%$& 99.22$\%$\\
  $\vert 00++\rangle$ & 0.0194 & 99.16$\%$& 98.95$\%$\\
	\hline
  \multicolumn{4}{|c|}{5 qubits} \\
	\hline
	$W_5$ & 0.1835 & 90.24$\%$& 75.89$\%$\\
	\hline
  \multicolumn{4}{|c|}{6 qubits} \\
	\hline
	Octahedron & 0.1234 & 83.23$\%$& 65.85$\%$\\
	$W_6$ & 0.1815 & 90.27$\%$& 75.28$\%$\\
	\hline
  \multicolumn{4}{|c|}{8 qubits} \\
	\hline
	Cube & 0.0890 & 70.93$\%$& 81.81$\%$\\
	$W_8$ & 0.1791 &  90.32$\%$& 75.06$\%$\\
  \hline
\end{tabular}
  \caption{$\mathcal{P}^n$ values of the pure states and corresponding fidelity thresholds for amplitude and phase damping noise.}
\end{table}

As we have observed previously, in general, the fidelity thresholds for phase damping are lower than the ones for amplitude damping, suggesting a greater robustness of most symmetric states under study to the former type of noise. This may be understood by the form of the Kraus operators, which are diagonal in the case of phase damping (see Eq. (\ref{eq:Krausphase})), thus affecting in a less significant way the density matrix of the state. An exception to this remark is the 8-qubit cube state; further investigation of other 8-qubit states might be helpful to elucidate this feature. We also observe that some states present a particularly pronounced difference between the two fidelity thresholds. In the case of the $W$ states, their small robustness to amplitude damping noise in terms of fidelity can be explained by the fact that $k=1$ in those states; indeed, photonic systems, for instance, are very sensitive to photon loss. It is therefore not surprising that the $S(4,2)$ state features a lower $F_{\text{amp,th}}$ value. The 6-qubit octahedron state also features an important difference between the two fidelity thresholds, indicating that, as in the 8-qubit case, it will be necessary to examine more such states in order to understand this property.

It is important to note that the threshold values shown in Tables II and III put stringent constraints on the experimental conditions required to observe non-locality in the presence of amplitude or phase damping. In photonics experiments, the detection efficiencies of Table II can be achieved using, for instance, superconducting transition edge sensors, for which the maximum reported efficiency reaches 95\% \cite{EFMP:rsi11}; however, note that losses here are attributed to inefficient detectors only while in practice more losses will occur in other parts of the setup too. Fidelities that can be achieved experimentally are around 80\% and 85\% for the $W_3$ and $S(4,2)$ state, respectively, in photonic systems \cite{MLF:prl05,KSTSW:prl07}, while values ranging from 85\% for the $W_4$ to 72\% for the $W_8$ state have been reported in trapped ion experiments \cite{HHR:nature05}. Comparing these values with the results shown in Table III allows identifying suitable realistic configurations for observing non-local correlations.

\section{Comparison between non-local tests}

\begin{figure}[tb]
      \centering
      \includegraphics[width=8cm]{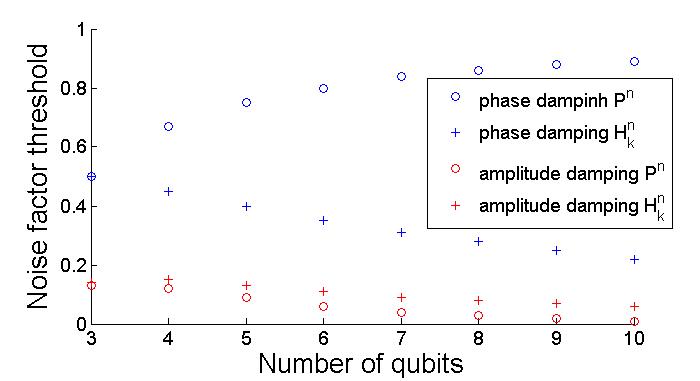}
      \caption{Noise factor thresholds corresponding to $\mathcal{P}^n$ (`o') and $\mathcal{H}^n_k$ (`+') inequalities as a function of the number of qubits in $W$ states under amplitude (red) and phase damping (blue) noise.}
			\label{fig:W_fid_hh}
\end{figure}

Our analysis up till now has focused on examining the robustness of several symmetric states based on the $\mathcal{P}^n$ Bell inequality of Eq. (\ref{eq:Pn}). It is now interesting to look into how these results may depend on the non-local test under study. To this end, we consider the Bell inequality $\mathcal{H}^n_k$ defined in Eq. (\ref{eq:Hnk}), and derive the noise factor thresholds, $\gamma_{\text{th}}$ and $\lambda_{\text{th}}$, in the case of amplitude and phase damping, respectively, for $W$ states, assuming a measurement strategy defined by the Majorana bases given in Section \ref{sec:W}. Our results for the thresholds as a function of the number of qubits in the $W$ states are shown in Fig. \ref{fig:W_fid_hh}.

We observe that the results given by the two tests in the amplitude damping case are very similar. The $\mathcal{H}^n_k$ test leads to slightly higher thresholds, which indicates a better robustness of this test to this type of noise, but in fact the situation is reversed for the $S(n,2)$ Dicke state. In the phase damping case, however, we observe a strikingly different behavior: the $\mathcal{H}^n_k$ test exhibits similar features as for amplitude damping, whereas the noise factor threshold tends to 1 for large numbers of qubits in the case of the $\mathcal{P}^n$ test. This has already been observed earlier, in Fig. \ref{fig:Wstate_ph}, and has been quantified in Eq. (\ref{eq:Wstate_th}). Clearly, the role of the non-locality test used to evaluate the robustness of a state to noise is of great importance and has to be carefully considered in each case.

\section{Effect of measurement strategy}\label{sec:measurement}

\begin{figure}[tb]
      \centering
      \includegraphics[width=8cm]{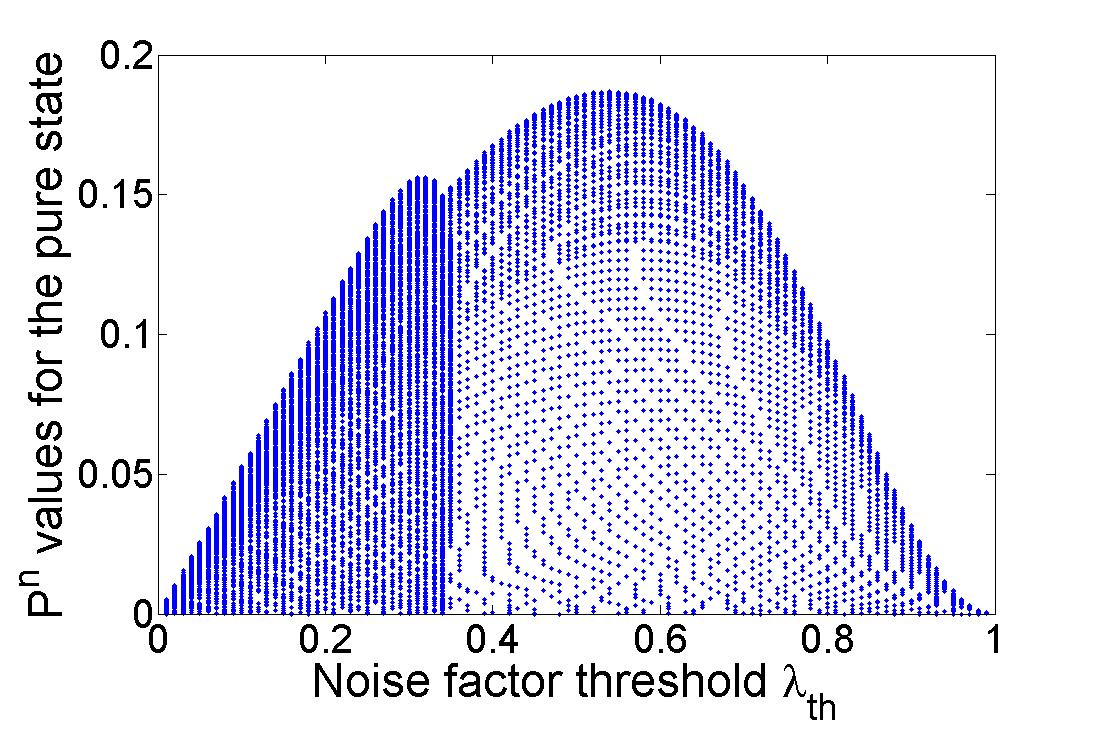}
      \caption{$\mathcal{P}^n$ values for the pure $W_4$ state vs phase damping noise factor threshold. Each point in this figure represents one particular pair of bases. The presence of two different peaks is due to the fact that there are two areas on the Bloch sphere that give rise to violation.}\label{fig:Distribution_of_Violation_and_Fidelity_Threshold}
\end{figure}

In the previous sections, we have used several measurement strategies to examine the robustness of the symmetric states under study to noise. We will now discuss the importance of the choice of this strategy focusing initially on the $W_4$ state and using the $\mathcal{P}^n$ test to derive the corresponding thresholds. In particular, in this case, we search numerically the entire Bloch sphere to identify the measurement strategies that optimize the noise factor threshold for phase damping, $\lambda_{\text{th}}$. The results are shown in Fig. \ref{fig:Distribution_of_Violation_and_Fidelity_Threshold}, where we plot the $\mathcal{P}^n$ values for the pure $W_4$ state versus the obtained values for $\lambda_{\text{th}}$ for a wide range of measurement settings.
In this figure, each point corresponds to a pair of measurement settings (bases). Horizontal lines indicate the measurement bases that give the same violation when no noise is considered but different noise factor thresholds, whereas vertical lines indicate the bases that give the same threshold but different violations for the pure state.
Obviously, the most interesting measurement strategies are those that give the maximum pure state $\mathcal{P}^n$ value for a given noise factor threshold, \emph{i.e.}, those that are situated at the envelope of the curve in Fig. \ref{fig:Distribution_of_Violation_and_Fidelity_Threshold}. Interestingly, some measurement strategies can still provide a non-zero violation even for a state under maximal phase damping noise. This comes however at the expense of very small violation levels, which indicates that the measurement strategies that are optimum in terms of violation are not necessarily optimum in terms of robustness. This leads to a trade-off that needs to be taken into account in a practical setting of quantum information protocol implementations.


\begin{figure}[tb]
      \centering
      \includegraphics[width=8cm]{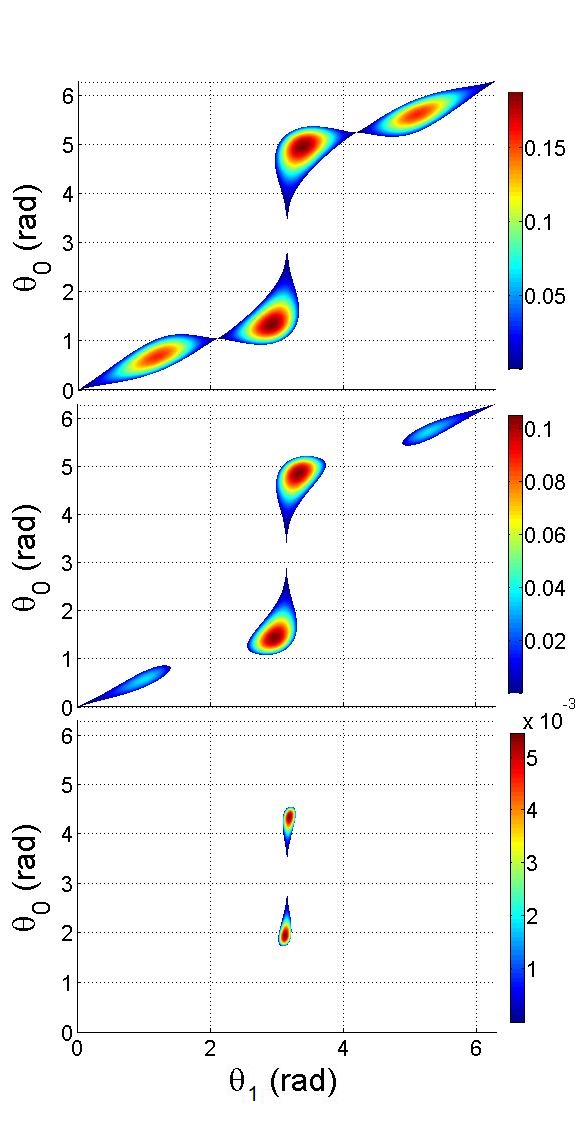}
      \caption{The colored surfaces represent the $\mathcal{P}^n$ values for the $W_4$ state under phase damping noise for noise factor $\lambda=0, 0.25, 0.77$ (from top to bottom). The angles $\theta_0$ and $\theta_1$ define the two measurement settings. We observe here the two different peaks that are also present in Fig. \ref{fig:Distribution_of_Violation_and_Fidelity_Threshold}.}
			\label{fig:W_2dfid_ph}
\end{figure}

To gain further intuition into this trade-off, we show in Fig. \ref{fig:W_2dfid_ph} a two-dimensional view of the $\mathcal{P}^n$ values for the $W_4$ state as a function of the measurement settings, for three different values of the phase damping noise factor, $\lambda = 0, 0.25, 0.77$. The measurement settings are defined here by the inclination angles, $\theta_0$ and $\theta_1$, while the difference between the azimuthal angles, $\varphi_0$ and $\varphi_1$, is set equal to $\pi$. In this figure, we observe a shift of the position of the maxima as well as a decreasing area of possible violation with increasing noise. Clearly, as the noise increases, the measurement settings that were giving the maximal violation for the pure state initially cannot give a violation above a certain threshold. Then, other possible measurement strategies may exist that can give a violation. This is true up to a global noise factor threshold, above which no violation can be obtained for any measurement setting. For the $W_4$ state, we find the maximum threshold $\lambda_{\text{th}}=0.996$. 

\begin{figure}[tb]
      \centering
      \includegraphics[width=8cm]{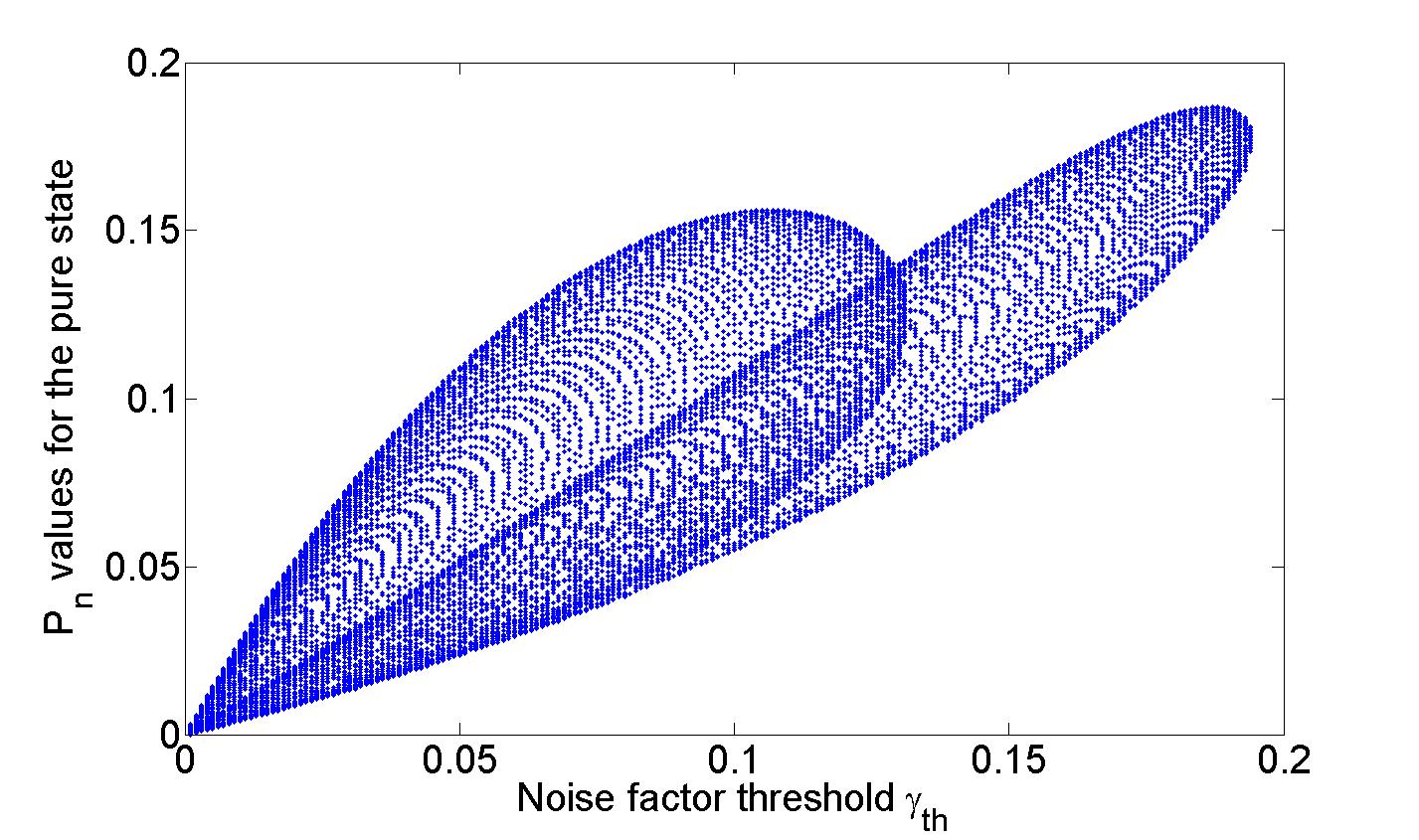}
      \caption{$\mathcal{P}^n$ values for the pure $W_4$ state vs amplitude damping noise factor threshold. Each point represents one particular pair of bases.}\label{fig:Distribution_of_Violation_and_Fidelity_Threshold_amp}
\end{figure}

Following the same procedure as for phase damping, we perform the optimization over the measurement strategies for the amplitude damping case, again for the $W_4$ state, seeking to maximize the corresponding noise factor threshold values, $\gamma_{\text{th}}$. The results are shown in Fig. \ref{fig:Distribution_of_Violation_and_Fidelity_Threshold_amp}. As we have seen earlier in Section \ref{sec:results}, the robustness of the symmetric states that we have examined differs significantly with the type of noise considered. It is therefore not surprising that the behavior with respect to the measurement strategy is quite different as well. Indeed, we observe here that the measurement settings that lead to high $\mathcal{P}^n$ values for the pure $W_4$ state are also the ones giving the optimal noise factor thresholds, contrary to the phase damping case, where good robustness can be obtained even at low violation levels. Hence, we conclude that in this case the optimum measurement strategy in terms of violation is also optimum in terms of robustness.

Finally, we can compare the two types of noise as well as two different measurement strategies, namely the Majorana and the optimum ones, in terms of the fidelity thresholds obtained in each case. In Table IV, we show these thresholds for the $W_{3,4,5,6}$ states, together with the $\mathcal{P}^n$ values for the corresponding pure states. We notice here that as expected from the previous discussion the Majorana basis, which gives smaller violations, leads to higher fidelity thresholds in the amplitude damping case. The values are also less spread than in the phase damping case, where in general the Majorana basis provides lower fidelity thresholds and hence better robustness.

\begin{table}[tb]
\center
\begin{tabular}{|l||c|c|c|c|}
   \hline
    State & $\mathcal{P}^n$ & $F_{\text{amp,th}}$ & $F_{\text{ph,th}}$ \\
   \hline
   \multicolumn{4}{|c|}{Majorana Basis} \\
   \hline
   $W_3$ & 0.1250 & 93.27$\%$ & 81.65$\%$ \\
   $W_4$ & 0.1250 & 93.81$\%$ & 70.53$\%$ \\
   $W_5$ & 0.0938 & 95.39$\%$ & 62.61$\%$ \\
   $W_6$ & 0.0625 & 96.95$\%$ & 57.74$\%$ \\
   \hline
   \multicolumn{4}{|c|}{Optimum Basis} \\
   \hline
   $W_3$ & 0.1926 & 90.04$\%$& 79.16$\%$\\
   $W_4$ & 0.1811 & 89.94$\%$& 77.14$\%$\\
   $W_5$ & 0.1835 & 90.24$\%$& 75.89$\%$\\
   $W_6$ & 0.1815 & 90.27$\%$& 75.28$\%$\\
   \hline
\end{tabular}
   \caption{$\mathcal{P}^n$ values of the pure state and fidelity thresholds for both types of noises and for two different measurement basis settings.}
\end{table}

\section{Sensitivity to alignment}

The measurement settings that determine the strategy followed by the parties in the non-local tests that we have considered are defined by precise choices of the corresponding inclination and azimuthal angles; these angle settings, however, have a limited precision in practice. For instance, in experiments where information is encoded in the polarization of photons, a typical error in the angles determining the desired polarization set by suitable waveplates is about $\pm2^{\circ}$, which translates to about $\pm0.07$ rad on the Bloch sphere. A natural question that arises then is the following: can such a small misalignment in the chosen angles affect significantly the possibility to observe non-locality under realistic conditions?

\begin{figure}[tb]
      \centering
      \includegraphics[width=8cm]{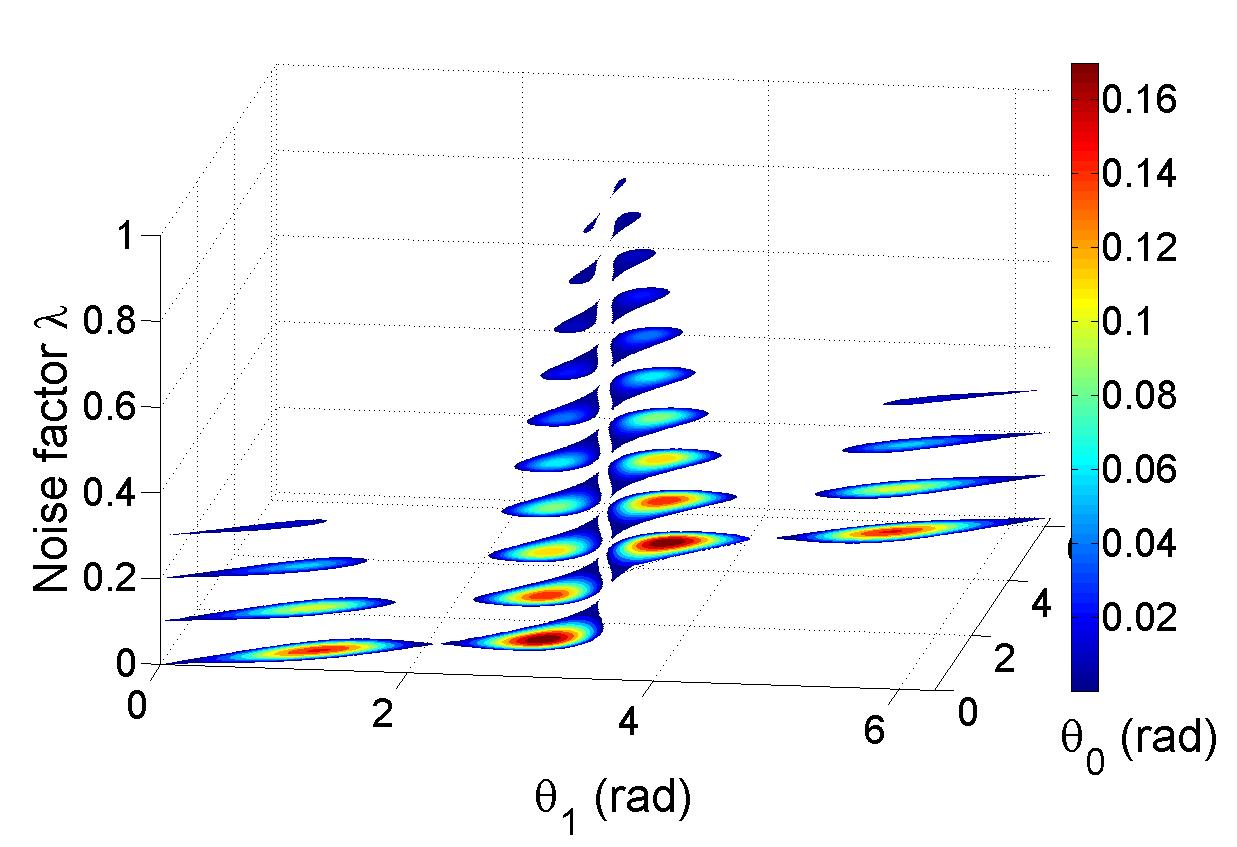}
      \caption{The colored surfaces represent the $\mathcal{P}^n$ values for the $W_4$ state under phase damping noise for several noise factor levels as a function of the inclination angles $\theta_0$ and $\theta_1$ that define the two measurement settings.}
			\label{fig:W_Fid_ph_t}
\end{figure}

In order to examine this question, we consider again the setting of Fig. \ref{fig:W_2dfid_ph}, for a larger range of noise factor values in Fig.~\ref{fig:W_Fid_ph_t}. This allows us to see how the surface of the violation evolves when the noise increases. The important point here is that the uncertainty in the angle settings may prohibit the observation of violation when the surface becomes small enough due to increased noise. Indeed, this uncertainty will in general lead to a lower noise factor threshold, which depends on the precision attainable with the available experimental equipment. Considering for example the aforementioned typical alignment error found in photonics experiments of $\pm2^{\circ}$, we find that the noise factor threshold decreases from $\lambda_{\text{th}} = 0.996$ to 0.81.

\begin{figure}[tb]
      \centering
      \includegraphics[width=8cm]{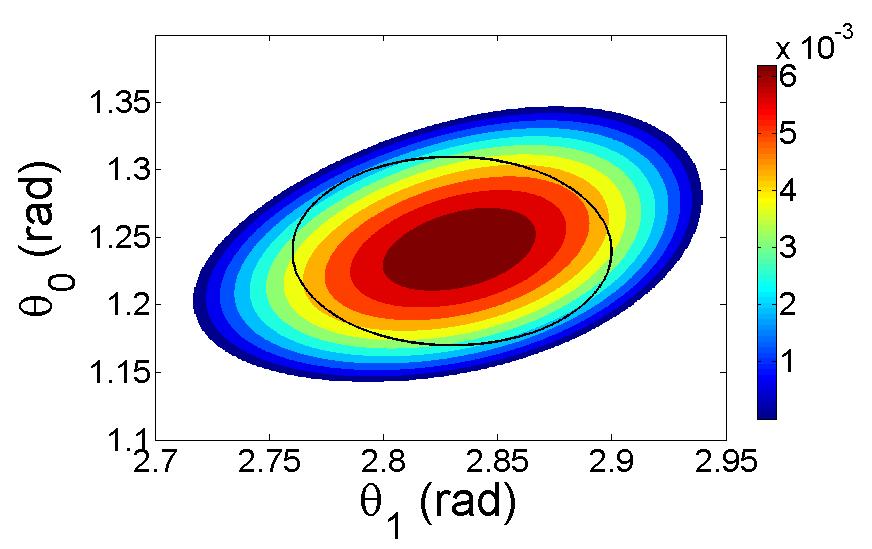}
      \caption{The colored surface represents the $\mathcal{P}^n$ values for the $W_4$ state under amplitude damping noise for noise factor $\lambda = 0.186$. The black oval line shows the typical angle sensitivity in photonics experiments and is centered around the maximal violation point.}
			\label{fig:Sensitivity}
\end{figure}

We perform the same analysis for the amplitude damping case, in the same setting, and we show in Fig. \ref{fig:Sensitivity} the results for a specific noise factor value. This allows us to observe in detail the variation of the violation value within the typical angle error surface for photonics experiments. Although for this noise factor the violation can be observed, it is clear that this will not be true for higher noise factors. Indeed, we find that in the amplitude damping case, the noise factor threshold decreases from $\gamma_{\text{th}} =0.2$ to 0.186 when the sensitivity to the angle setting is taken into account.

\section{Application: Degeneracy class discrimination using non-local tests}

As an application of the techniques that we have developed in the previous sections, we would like to consider the case of symmetric states exhibiting degeneracy $d$, as discussed in Section \ref{sec:states}. In particular, we are interested in determining the conditions under which the non-locality of such states can be observed in the presence of noise but also whether states featuring different degeneracy can be discriminated in such conditions. The available tool for this purpose is the Bell inequality $\mathcal{Q}^n_d$, defined in Eq. (\ref{eq:Qnd}), which has been shown to be violated by any (pure) Dicke state with degeneracy $d$ \cite{WM:prl12}. It is important to note here that the discrimination of entanglement classes using a non-local test \cite{WM:prl12,WM:pra13,BSV:prl12} implies that this task can be performed in a \emph{device independent way}; this means that even when the parties do not have control over their measurement equipment, observing a violation of the $\mathcal{Q}^n_d$ inequality, for instance, proves that the state under study belongs to the corresponding degeneracy class.

We perform this analysis for three 6-qubit Dicke states featuring different degeneracies and we find the results summarized in Table V. We can first observe that if we obtain the result $\mathcal{Q}^6_4 > 0$ (with a maximum at $\mathcal{Q}^6_4 = 0.0177$ reached by the $S(6,1)$ state using the measurement strategy given by $\mathcal{M}_0 = \{\theta_0=0.60$, $\varphi_0=0\}$ and $\mathcal{M}_1 = \{\theta_1=0.65$, $\varphi_1=\pi\}$), then the state under study is necessarily $S(6,1)$, regardless of the measurements performed on the other states. This result therefore identifies the class of Dicke states with degeneracy $d > 4$, in a device independent way. Furthermore, we can see that if we restrict our analysis to Dicke states with degeneracy $d = 2$ or 3, then the result $\mathcal{Q}^6_3 > 0$ identifies the state $S(6,2)$  (with a maximum at $\mathcal{Q}^6_3 = 0.0069$ obtained using the measurement strategy given by $\mathcal{M}_0 = \{\theta_0=0.56$, $\varphi_0=0\}$ and $\mathcal{M}_1 = \{\theta_1=0.61$, $\varphi_1=\pi\}$).


\begin{table}[h!]
\center
\begin{tabular}{|>{\centering}m{0.15\linewidth}||>{\centering}m{0.25\linewidth}|>{\centering}m{0.25\linewidth}|>{\centering}m{0.25\linewidth} | >{\centering}m{0.2\linewidth}|}
	\hline
	Test & $\mathcal{Q}^6_3$ & $\mathcal{Q}^6_4$ & $\mathcal{Q}^6_5$ \tabularnewline \hline \hline \vspace{0.1cm}
	$S(6,1)$ & $0.0519$ & $0.0177$ & 0  \tabularnewline \hline \vspace{0.1cm}
	$S(6,2)$ & $0.0069$ & 0 & X \tabularnewline \hline \vspace{0.1cm}
	$S(6,3)$ &  0 & X & X   \tabularnewline \hline
\end{tabular}
\caption{$\mathcal{Q}^6_d$ violations for three 6-qubit Dicke states. The symbol X means that there is no possible violation.}
\end{table}

Applying the techniques of the previous sections, we can further determine conditions under which this degeneracy class discrimination is possible in the presence of phase damping noise. In particular, we find that, for the $S(6,1)$ state, a violation can be observed when the noise factor is below the value $\lambda_{\text{th}} = 0.004$, while an angle precision of $\pm0.15^{\circ}$ is required in this case, assuming a photonics implementation using polarization encoding. Finally, the $S(6,2)$ state can be discriminated for a noise factor below $\lambda_{\text{th}} = 0.008$, with a required precision of $\pm0.1^{\circ}$. Clearly, these thresholds put stringent constraints in experiments aiming at demonstrating such non-local properties in practice.

\section*{CONCLUSION}

In this work, we have performed a detailed study of the robustness of the non-locality exhibited by a wide range of permutation symmetric states against decoherence in the form of amplitude and phase damping. Our analysis was based primarily on a Bell inequality using an extended version of Hardy's paradox to test the non-locality of symmetric states. This work is motivated by the need to develop experimentally relevant criteria for observing non-local correlations that are useful for quantum information applications in the complex multipartite case. To this end, we performed a full optimization analysis over the measurement settings used for the non-local tests, and have thus derived the levels of violation that can be achieved in realistic conditions as well as thresholds for the observation of non-locality, quantified by several parameters of importance in experimental implementations. Although the obtained values set stringent constraints for such implementations, it is possible to perform some of the tests that we have considered with current technology.

For the $W$ states, we have found that the fidelity threshold for phase damping noise scales as $1/\sqrt{n}$ with the number of qubits, $n$, in the state; our result, derived using a Bell inequality based on the full range of multipartite correlations, confirms previous results on the robustness of the non-locality of these states against dephasing, derived using CHSH-type correlations \cite{CAAC:pra}, and indicates these states as being particularly well suited for practical quantum information applications involving many qubits. For these states, and for several others, we have additionally observed the different behavior with respect to the type of noise under consideration. For example the octahedron state seems more robust against amplitude damping than the $W$ or Dicke states.

We have demonstrated the importance of the measurement basis choice, of practical imperfections such as the limited precision of standard components, and, for the phase damping case, of the choice of the non-local test. In general, it is important to continue the search for Bell inequalities tailored to specific states that can provide better thresholds with correspondingly higher violations. For instance, in the discrimination case that we have examined, very low noise thresholds are obtained; this is mainly due to the fact that the $\mathcal{Q}^n_d$ inequality is a subclass of the general $\mathcal{P}^n$ inequality, hence leading to less tight results, albeit with the possibility to detect different degeneracy classes.

Finally, a further analysis should consider the case where each party chooses a measurement basis for the non-local test independently. This will require the use of advanced optimization tools such as semidefinite programming. We expect that such extensions of our techniques will be important for studying the practical implementation of a wide range of device independent quantum information tasks.\\

\noindent {\bf Acknowledgements.} We thank Nicolas Treps for useful comments. This work was supported by the Ville de Paris Emergences project CiQWii.

\bibliography{decoherence_nonlocality}

\end{document}